\documentclass[usegraphicx]{mn2e}

\usepackage{amssymb}
\usepackage{mathptmx}
\usepackage{url}

\newcommand{\kpc}{\rm\thinspace kpc}

\newcommand{\km}{\rm\thinspace km}

\newcommand{\cm}{\rm\thinspace cm}
\newcommand{\yr}{\rm\thinspace yr}

\newcommand{\s}{\rm\thinspace s}

\newcommand{\Msun}{\hbox{$\rm\thinspace M_{\odot}$}}

\newcommand{\Msunpyr}{\hbox{$\Msun\yr^{-1}\,$}}

\newcommand{\keV}{\rm\thinspace keV}

\newcommand{\kmps}{\hbox{$\km\s^{-1}\,$}}

\newcommand{\Zsun}{\hbox{$\thinspace \mathrm{Z}_{\odot}$}}

\newcommand{\pcm}{\hbox{$\cm^{-3}\,$}}
\newcommand{\psqcm}{\hbox{$\cm^{-2}\,$}}
\newcommand{\pcmsq}{\hbox{$\cm^{-2}\,$}}

\begin{document}

\title{Spatially-resolved X-ray spectroscopy of the core of the
  Centaurus cluster}
  \author[J.S. Sanders \& A.C. Fabian]
  {J.S. Sanders\thanks{E-mail: jss@ast.cam.ac.uk} and A.C. Fabian\\
  Institute of Astronomy, Madingley Road, Cambridge. CB3 0HA}
\maketitle

\begin{abstract}
  We present \emph{Chandra} data from a 31.7~ks observation of the
  Centaurus cluster, using the ACIS-S detector. Images of the X-ray
  emission show a plume-like feature at the centre of the cluster, of
  extent 60~arcsec (20~kpc in projection). The feature has the same
  metallicity as gas at a similar radius, but is cooler.  Using
  adaptive binning, we generate temperature, abundance and absorption
  maps of the cluster core.  The radial abundance profile shows that
  the previously known, steep abundance gradient peaks with a
  metallicity of $1.3-1.8 \Zsun$ at a radius of about 45~arcsec
  (15~kpc), before falling back to $0.4\Zsun$ at the centre of the
  cluster. A radial temperature profile shows that the temperature
  decreases inwards. We determine the spatial distributions of each of
  two temperature components, where applicable.  The radiative cooling
  time of the cooler component within the inner 10~arcsec (3~kpc) is
  less than $2\times 10^7$~yr. X-ray holes in the image coincident
  with the radio lobes are seen, as well as two outer sharp
  temperature drops, or cold fronts. The origin of the plume is
  unclear. The existence of the strong abundance gradient is a strong
  constraint on extensive convection or gas motion driven by a central
  radio source.
\end{abstract}


\section{Introduction}
The Centaurus cluster, Abell~3526, at a redshift $z=0.0104$ (Lucey,
Currie \& Dickens 1986) is a nearby, X-ray bright galaxy cluster.
Analysis of \emph{ASCA}, and earlier data (e.g. Allen et al. 2001)
indicates it hosts a cooling flow (see Fabian 1994 for a review) of
30-50~\Msunpyr, centred on the cD galaxy NGC~4696.  Previous
observations of Centaurus (e.g. Allen \& Fabian 1994) show a system
with smooth, elliptical, X-ray isophotes, indicating the system is
relatively relaxed. However, there is evidence for a current or past
merger event (Allen \& Fabian 1994; Churazov et al. 1999; Furusho et
al. 2001). The neighbouring subcluster, Cen~45, centred on NGC~4709,
has a velocity which is 1500~\kmps higher than the main Centaurus
cluster, also known as Cen~30 (Lucey, Currie \& Dickens 1986).

Observations of the cluster using \emph{ROSAT} and \emph{ASCA} show
that the central region of the cluster is rich in metals, with a large
abundance gradient (Fukazawa et al. 1994; Ikebe et al. 1998; Allen et
al. 2001).

\section{The X-ray image}
The data presented here are from a 31.7~ks observation of the
Centaurus cluster by the \emph{Chandra} Observatory, using the ACIS-S
detector. The temperature of the focal plane was $-120^{\circ}$C at
the time of the observation. No period in the observation was affected
by flares, so no data were excluded from the analysis. The interaction
of the radio source PKS~1246-410 with the X-ray plasma is discussed
elsewhere (Taylor, Fabian \& Allen 2001).  All errors quoted are
1-$\sigma$ uncertainties.  Distances assume $H_0 = 50 \textrm{ km
  s}^{-1} \: \textrm{Mpc}^{-1}$.  Data were fitted using the
\textsc{xspec} package (version 11.0.1; Arnaud 1996) and metallicities
are relative to solar abundances defined by Anders \& Grevesse (1989).

\begin{figure}
  \begin{center}
    \includegraphics[width=0.99\columnwidth]{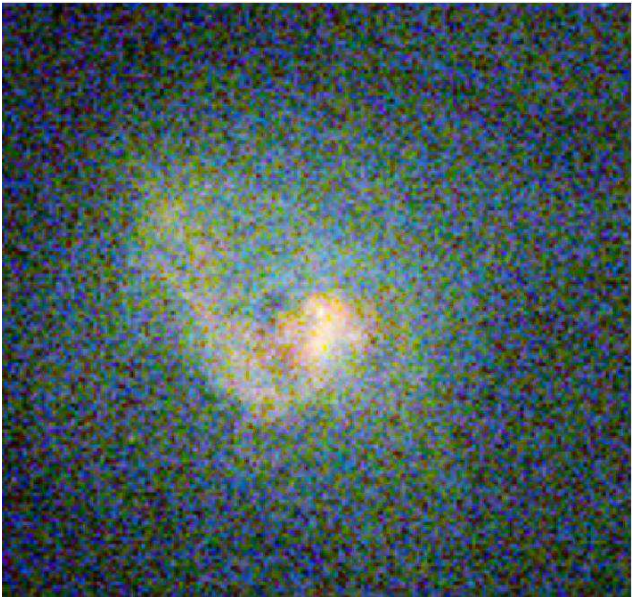}
  \end{center}
  \caption{Colour X-ray image of the inner part of the ACIS-S3
    chip centred on the core of the Centaurus cluster. Events between
    0.5 and 1~keV are coloured red, those between 1 and 2~keV green,
    and those between 2 and 7~keV blue. The pixel size is $\sim 1$~arcsec.
    The image measures $\sim 140$~arcsec (46~kpc) vertically. The
    emission peaks at around 110 counts per square-arcsecond between
    0.5 and 7~keV.}
  \label{fig:imageRaw}
\end{figure}

Fig. \ref{fig:imageRaw} shows a coloured raw ($\sim 1$~arcsec pixels)
image of the inner 140~arcsec of the S3 chip. Events are coloured
according to the energy recorded (0.5-1~keV red, 1-2~keV green,
2-7~keV blue).  The plume-like structure to the NE resembles a twisted
sheet, and was seen in an observation of the cluster using the
\emph{ROSAT} HRI (Sparks, Jedrzejewski \& Macchetto 1993).

\begin{figure}
  \begin{center}
    \includegraphics[width=0.99\columnwidth]{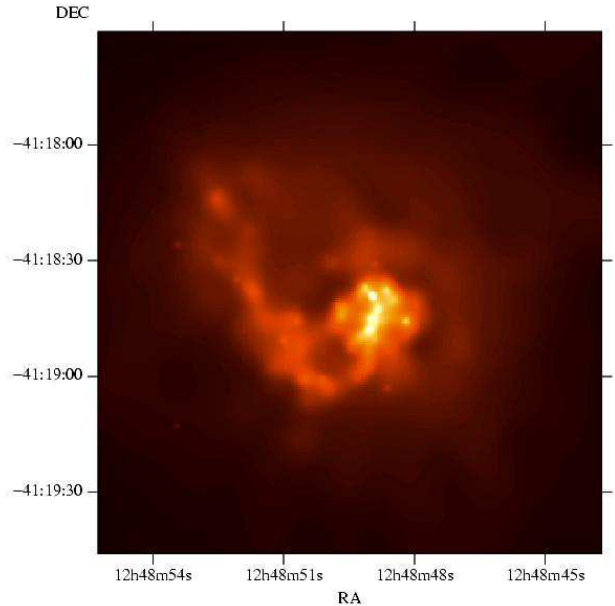}
  \end{center}
  \caption{Adaptively smoothed image of the centre of the Centaurus
    cluster in the energy range 0.5 to 7~keV with $\sim 0.5$~arcsec
    pixels. The image measures 175 arcsec vertically, which
    corresponds to a distance of 58~kpc.  The minimum \textsc{asmooth}
    significance is 3-$\sigma$.}
  \label{fig:imageASmooth}
\end{figure}

Fig. \ref{fig:imageASmooth} shows an adaptively-smoothed (Ebeling,
White \& Rangarajan 2001) image of the raw data. It has been smoothed
with a minimum significance of 3-$\sigma$. The image highlights that
the emission is diminished from the north-east to the south-west close
to the nucleus. It clearly shows the spiral plume-like feature,
appearing to emerge from the nucleus in the south, and heading towards
the north-east. The plume is approximately 60~arcsec long (20~kpc in
projection). The energy-coloured image (Fig. \ref{fig:imageRaw}) shows
the centre of the cluster, and the plume, is cooler (redder) than the
rest of the cluster. The two deficits or `holes' in emission
immediately to the NE and NW of the core are associated with the lobes
of the radio source (Taylor et al. 2001). The continuation of the
holes to the SE and SW is probably also related with the southern
extension to the radio lobes.

Fabian et al. (2001b) hypothesise that the filament-like feature seen
in X-ray emission in the core of the Abell~1795 cluster may be caused
by the relative motion of the cD galaxy to the rest of the cluster
(estimated to be at least 150~\kmps). The results of Dickens, Currie
\& Lucey (1986) show that NGC~4696 lies close to the mean velocity of
Cen~30. However, the radio maps of Taylor et al. (2001) show that the
radio lobes bend to the south of the core. This indicates that
NGC~4696 is moving north relative to the local intracluster medium.
The aspect of the radio lobes could include a projection effect, with
a velocity component along our line of sight of 100-200\kmps.

\begin{figure}
  \begin{center}
    \includegraphics[width=0.99\columnwidth]{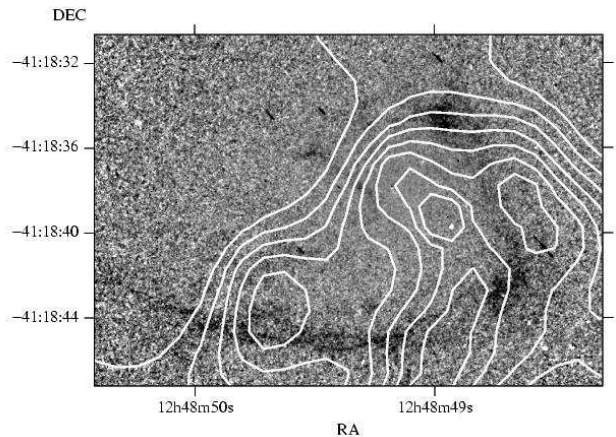}
  \end{center}
  \caption{Optical \emph{HST} image showing the dust lane in the centre of
    NGC~4696 overlayed by contours from the adaptively smoothed image
    of the X-ray data (Fig.~\ref{fig:imageASmooth}). The optical image
    shows the ratio of the emission in a band with a central
    wavelength of 5407~\r{A} relative to one at 7940~\r{A}.}
  \label{fig:imageHST}
\end{figure}

Observations of the core of the cluster at optical wavelengths show a
dust-lane in NGC~4696 (Shobbrook 1963). It has been noted that the
X-ray emission has a similar morphology to the dust lane (Sparks et
al. 1993).  Using four \emph{HST} archived observations (datasets
U3560101T, U3560102T, U62G8401R and U62G8402R) of the galaxy at
5407~\AA{} and 7940~\AA, we have made a ratio image of the optical
emission in two bands highlighting the dust-lane. This is shown in
Fig.~\ref{fig:imageHST}, overlayed with a contour map of the centre of
the adaptively smoothed image, Fig.~\ref{fig:imageASmooth}. The
swirling morphology of the dust-lane is similar to, but smaller than,
the X-ray plume.

We find no evidence for hard X-ray emission (between 2-7 or 4-7~keV)
from the core of the radio source PKS~1246-410 (position taken from
Taylor et al. 2001) at the 2-$\sigma$ level.  Hard X-ray emission was
seen in Centaurus using \emph{ASCA} (Allen, Di Matteo \& Fabian 2000),
which has a hundredfold larger point-spread function than
\emph{Chandra}.  There is tentative evidence for diffuse hard X-ray
emission in the \emph{Chandra} data which will be studied elsewhere.

\section{Spectral fits}
\subsection{Spectrum of central 10 arcsec}
\label{sect:spec_centre}
We extracted a spectrum from the central 10 arcsec radius of the
cluster. The spectrum was fitted with an absorbed single-temperature
(1-T) \textsc{mekal} model (Mewe, Gronenschild \& van den Oord 1985;
Liedahl, Osterheld \& Goldstein 1995), in the energy range of 0.5 to
7~keV, where the \emph{Chandra} response-matrix is well calibrated.
The spectrum was binned to ensure a minimum number of counts per
spectral bin, chosen here to be 22 arbitrarily.  We generated an
effective response and ancillary matrix, averaged from the matrices
applicable to each part of the region, but weighted according to the
relative number of counts. The background was removed using a blank
field background-spectrum of the same part of the detector, created
using the prescription of Markevitch (2001).  Although the background
fields have lower galactic absorption ($N_{\mathrm{H}} = 0.02 \times
10^{22}\pcmsq$) than the region observed ($0.081 \times 10^{22}
\pcmsq$, obtained using the \textsc{ftools} \textsc{nh} program;
Dickey \& Lockman 1990) its level is insignificant compared with the
bright cluster emission studied here.

The results of the fit are summarised in Table \ref{tab:regionfits}.
We find a temperature of 0.71~keV, a metallicity of $0.16\Zsun$,
absorbed by $N_{\mathrm{H}} = 0.26\times10^{22} \pcmsq$.  The 1-T
model poorly fits the spectrum (Reduced $\chi^2 = 660/128 = 5.2$).

In regions of cooling flow clusters where radiative cooling dominates,
i.e. where the radiative cooling time is shorter than the age of the
flow (about 3--5 Gyr; Allen \& Fabian 1997, corresponding to about one
arcmin radius in the Centaurus cluster), the X-ray surface brightness
profiles indicated, if there is no balancing heat source, that the
inferred mass deposition of cooled gas is roughly proportional to
radius ($\dot{M} \propto r$). This requires that the gas there is
multiphase (Nulsen 1986; Thomas, Fabian \& Nulsen 1987). The spectrum
of multiphase gas should be fit better with a two-temperature (2-T)
model than a 1-T model (e.g. Allen et al. 2001).  An even better model
should be a cooling flow (CF: Johnstone et al. 1992), possibly with
intrinsic absorption.

Studies of several cooling flow clusters using the \emph{XMM-Newton}
RGS (e.g. Peterson et al. 2001; Tamura et al. 2001) have shown that
soft X-ray emission lines expected from gas cooling to low
temperatures are absent in the data. A model (cooling flow to a
minimum temperature, CFMT) which does fit the data, is a cooling flow
in which gas cools from the upper temperature to a lower temperature,
which may be one third to one fifth of the upper one. Such a model
requires that the gas disappears once it reaches the minimum
temperature, perhaps because of intrinsic absorption for which the
absorbed CF model is an approximation, because of mixing or because of
rapid heating which turns the cool phase into the hot one (see Fabian
et al.  2001a).

In this paper we try all these models (1-T, 2-T, CF and CFMT) on the
data. Previous work with \emph{ASCA} has shown that more than one
temperature component is required in the core of the Centaurus cluster
(Fukazawa et al. 1994; Ikebe et al. 1998; Allen et al. 2001).
Projection effects due to hotter outer gas being along the line of
sight to cooler gas are also examined for the plume. We do not however
carry out an exhaustive set of modelling due to the unknown
3-dimensional geometry of the cooler gas and the limitations of the
models themselves, which do not include effects due for example to
gravitational heating in a flow. Moreover, as we show, the metallicity
of the gas is spatially complex, which further complicates detailed
modelling. It is our intention here to identify the gross behaviour of
the gas in terms of its temperature and abundance distributions with
position.

We fitted the spectrum of the core with a 2-T \textsc{mekal} model,
fixing the abundances of the two phases to be the same as there is not
enough information in the spectrum to measure the abundances
separately and accurately. The results of this fit are also shown in
Table \ref{tab:regionfits}. The temperatures of the two phases are
around 0.7~keV and 1.4~keV. The metallicity ($\sim 0.6 \Zsun$) is much
larger than for the 1-T result ($\sim 0.2 \Zsun$). The quality of the
fit is considerably better than the 1-T model (Reduced $\chi^2 =
200/132 = 2.4$), but still not good. Allowing the two abundances to
vary did not significantly improve the fit. The best-fitting
metallicities were $1.0_{-0.4}^{+1.7}\Zsun$ (cold-phase) and
$0.46_{-0.06}^{+0.08}\Zsun$ (hot-phase), but the temperatures were
unaffected. Adding a third temperature to the 2-T model did not
significantly improve the fit.  However, adding a third component
fixed at 4~keV to account for projected emission was an improvement.
The radiative cooling time of the cooler component in the 2-T fit is
about $2\times 10^7\yr$.

The fits can be improved further by switching from a \textsc{mekal}
model to an \textsc{apec} model (Brickhouse et al. 2000). The reduced
$\chi^2$ for an \textsc{apec} 2-T fit is 2.0.  The residuals of the
fit show there is some excess emission above 3~keV.  This emission can
be fit either using a power-law or a high-temperature plasma. We show
in Table \ref{tab:regionfits} the slight changes induced by adding a
plasma of fixed temperature 4~keV and metallicity $0.4\Zsun$. Adding
this component is justified as we expect contributing emission from
the cluster in front of the core.

The spectrum can also be fitted (but with a poorer goodness-of-fit)
using an intrinsically absorbed cooling flow model (\textsc{rjcool};
Johnstone et al. 1992), plus a plasma fixed at 4~keV and $0.4\Zsun$.
We absorbed both components by a fixed galactic absorber, and allowed
the intrinsic absorption of the cooling flow to vary. Normally we
would add another plasma component to account for the local emission
of the cluster near the cooling gas, and gravitational work done.
However, at the centre of the cluster the cooling flow dominates, so
no extra component was necessary.

\begin{figure}
  \begin{center}
    \includegraphics[angle=-90,width=0.99\columnwidth]{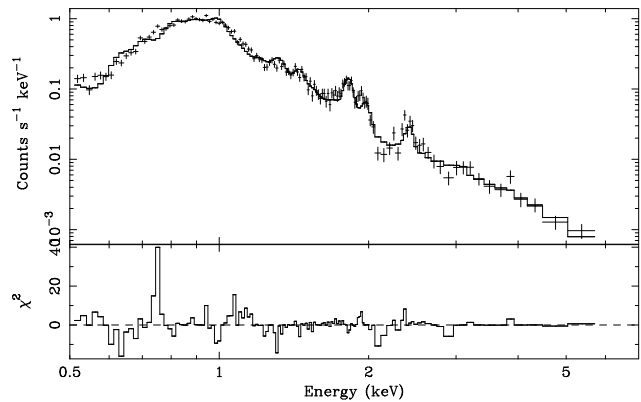}
  \end{center}
  \caption{Spectrum of the inner 10 arcseconds of the cluster, fitted
    with a 1-T \textsc{mekal} model (fixed at 4~keV and $0.4\Zsun$),
    plus an intrinsically absorbed \textsc{rjcool} cooling flow.}
  \label{fig:speccore}
\end{figure}

We derive a cooling flow mass deposition rate of $11 \Msunpyr$. The gas
cools from a temperature of 1.1~keV and has a metallicity of $0.7
\Zsun$.  The cooling flow is intrinsically absorbed by $2.4 \times
10^{21} \pcmsq$. The spectrum, fitted with the cooling flow model, is
shown in Fig.  \ref{fig:speccore}.  The cooling flow model calculates
its emission using \textsc{mekal} plasmas. We speculate that an
\textsc{apec} version of the code would fit the spectra better, but
unfortunately this code is still under development.

Using a CFMT model (Table~\ref{tab:regionfits}) we find the mass
deposition rate and upper temperature to be unchanged, a decrease in
the intrinsic absorption of the cooling flow to a value similar to
that found using 2-T fits, and a lower cooling temperature of
0.40~keV. The fit is better than a single cooling flow, but not as
good as a 2-T model.

All the fits we have tried to the core spectrum show a large residual
around 0.75~keV, independent of the model used (shown for the
\textsc{rjcool} model in the lower panel of Fig. \ref{fig:speccore}).
\textsc{mekal} and \textsc{apec} models with varying elemental
abundances do not improve the fit to this feature.  The emission is
well fitted with a $\delta$-function. By examining the spectra of
annuli, we have determined that this feature is not significant
outside a radius of 10~arcsec (3~kpc) from the centre. It is also
present in the inner 5~arcsec. The effective area of the telescope
varies smoothly over the energy range of the emission, making the
feature unlikely to be a calibration error.

There are also negative-residuals just above 2~keV.  The effective
area per unit energy of the telescope almost halves in the range 2 to
2.1~keV, so this feature may be a calibration issue. Accounting for
these two spectral features would greatly improve the fit to the data,
and would lead to a much smaller reduced $\chi^2$.

\begin{figure}
  \begin{center}
    \includegraphics[angle=-90,width=0.85\columnwidth]{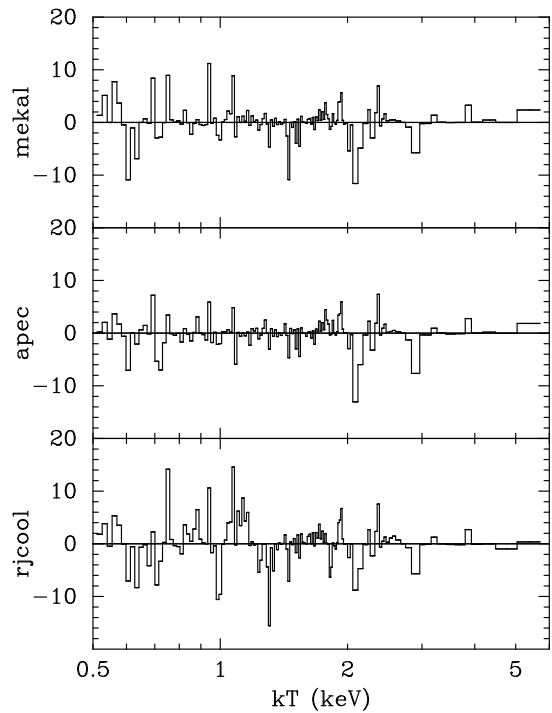}
  \end{center}
  \caption{Contribution to $\chi^2$ (times sign of the difference
    between model and spectrum) of fitting the spectrum of the inner
    10~arcsec with three different models, including the contribution
    of the emission at 0.75~keV. (top) 2-T \textsc{mekal} model plus a
    fixed component at 4~keV and $0.4\Zsun$. (middle) 2-T
    \textsc{apec} model with a fixed component. (bottom)
    \textsc{rjcool} component plus another fixed component.}
  \label{fig:residuals}
\end{figure}

We show in Fig. \ref{fig:residuals} the contributions towards $\chi^2$
for three different fits to the spectrum of the core. The feature at
0.75~keV, in all spectra, has been removed by fitting it with a
Gaussian. We also included a plasma fixed a 4~keV and $0.4\Zsun$ to
model the projected cluster emission. The top panel shows a 2-T
\textsc{mekal} model, the middle panel a 2-T \textsc{apec} model, and
the lower panel an intrinsically absorbed \textsc{rjcool} model with
fixed galactic absorption.

\subsection{Spectrum of the plume}
\label{sect:spec_plume}
We extracted a spectrum from the plume-like region in the cluster. It,
like the core, is poorly fit by a 1-T model. Using a 2-T model, the
fitted temperature of the hot component is $\sim 0.1\keV$ larger than
the hot component in the core. The cooler component has a similar
temperature. The abundance of this region is significantly higher than
the core (an increase of $0.2-0.3 \Zsun$).

\begin{figure}
  \begin{center}
    \includegraphics[angle=-90,width=0.99\columnwidth]{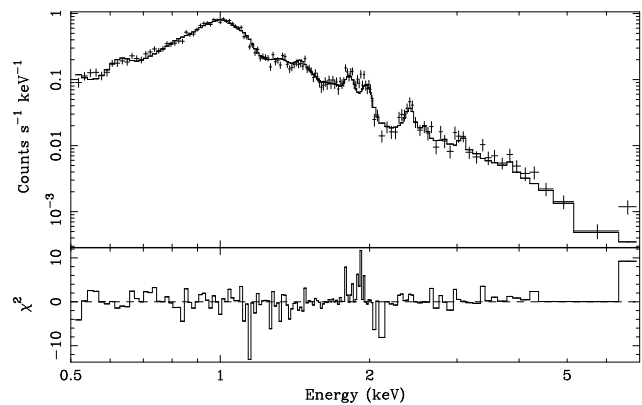}
  \end{center}
  \caption{Spectrum of the plume, fitted with a 2-T
    \textsc{apec} model.}
  \label{fig:specplume}
\end{figure}

We also fitted the spectrum of an area at a similar radius to the
plume (but not including the plume) with a 2-T model. The results are
shown in Table \ref{fig:specplume} as `plume radius'. The plume
appears to be $\sim 0.25$~keV cooler (in both fitted temperatures).
There is no significant abundance difference.  A cooling flow model
(with a component to account for local cluster emission) gives a mass
deposition rate of $4 \Msunpyr$, with a high metallicity of
$1.6\Zsun$.

Our next task was to test whether the multi-temperature nature of the
plume is a projection effect. We compared the emission measure of the
upper temperature component of a 2-T model fitted to the plume, with
the emission measure of a 1-T fit to the spectrum in another region
with the same size and radius as the plume, but at an angle of
$90^{\circ}$ to it. The emission measure of the upper-temperature
component of the plume was 70~percent greater than the emission
measure of this comparison region.  Therefore, the multi-temperature
nature of the plume is not wholly a projection effect but is intrinsic
to the plume.

By assuming a cylindrical geometry for the plume (20~kpc long, radius
2.3~kpc), and fitting a 1-T model to its spectrum, the emission
measure implies an electron density of $\sim 0.13$~\pcm. The total gas
mass of the plume is therefore $\sim 10^9$~\Msun. This mass is
compatible with the hypothesis that this gas was stripped off an
infalling galaxy. The mean cooling time of the gas in the plume is
approximately $10^8$~yr, so the gas, or galaxy, must have a velocity
of $\gtrsim 200$~\kmps in order to not have cooled. The similarity of
the abundance of the plume to the other gas at that radius raises a
problem for the stripping hypothesis.

\begin{table*}
  \begin{center}

    \begin{tabular}{llllllll}
      Area & Counts & Model & $N_{\mathrm{H}}$
      & $kT_1$ (keV) & $kT_2$
      (keV) & $Z$ (\Zsun) & Reduced $\chi^2$ \\ 
      &&& $(10^{22} \pcmsq)$ && or $\dot{M}$ ($\Msunpyr$) \\
      \hline

      Central 10 arcsec & 17835 &
      phabs(mekal) &
      $0.26 \pm 0.01$ & $0.71 \pm 0.01$ & & $0.16 \pm 0.01$ &
      $660/128 = 5.2$ \\

      && phabs(apec) &
      $0.23 \pm 0.01$ & $0.76 \pm 0.01$ & & $0.14 \pm 0.01$ &
      $671/128 = 5.2$ \\

      && phabs(mekal + mekal) &
      $0.19 \pm 0.01$ & $0.67 \pm 0.01$ & $1.41 \pm 0.05$ &
      $0.59^{+0.07} _{-0.05}$ & $299/126 = 2.4$ \\

      && phabs(apec + apec) &
      $0.16 \pm 0.01$ & $0.68 \pm 0.01$ & $1.45 \pm 0.03$ &
      $0.62 \pm 0.06$ & $256/126 = 2.0$ \\

      && phabs(apec+apec+[apec])${}^{*}$ &
      $0.16 \pm 0.01$ & $0.66 \pm 0.01$ & $1.33 \pm 0.03$ &
      $0.64_{-0.05}^{+0.08}$ & $243/125 = 1.9$ \\

      && phabs(zphabs(rjcool) + &
      $0.24 \pm 0.02$ & $1.14 \pm 0.02$ & $\dot{M} = 11.3 \pm 0.7$ &
      $0.71^{+0.11}_{-0.08}$ & $371/127=2.9$ \\
      && \hfill [mekal])${}^\dagger$ \\

      && phabs(zphabs(rjcool &
      $0.16 \pm 0.02$ & $1.15 \pm 0.03$ & $\dot{M} = 11.0 \pm 0.7$ &
      $0.49^{+0.05}_{-0.05}$ & $339/126=2.7$ \\
      && \hfill - rjcool) + [mekal])${}^\ddagger$
      && $0.40 \pm 0.02$ \\
      \\
      Plume & 14206 &
      phabs(mekal) &
      $0.18 \pm 0.01$ & $1.14 \pm 0.01$ & & $0.28 \pm 0.01$ &
      $583/138 = 4.2$ \\

      && phabs(apec) &
      $0.17 \pm 0.01$ & $1.22 \pm 0.01$ & & $0.30 \pm 0.02$ &
      $451/138 = 3.3$ \\

      && phabs(mekal + mekal) &
      $0.17 \pm 0.01$ & $0.79 \pm 0.02$ & $1.58 \pm 0.04$ &
      $0.89^{+0.10}_{-0.08}$ & $278/136 = 2.0$ \\

      && phabs(apec + apec) &
      $0.16 \pm 0.01$ & $0.82 \pm 0.02$ & $1.62 \pm 0.03$ &
      $0.81 \pm 0.08$ & $219/136 = 1.6$ \\

      && phabs(mekal + [mekal] + &
      & $1.56 \pm 0.06$ & & $1.58^{+0.35}_{-0.23}$ &
      $289/136 = 2.1$ \\
      && \hfill zphabs(rjcool))${}^\S$ &
      $0.28 \pm 0.02$ & &
      $\dot{M} = 4.1 \pm 0.3$ \\

      && phabs(mekal + [mekal] + &
      & $1.94^{+0.26}_{-0.18}$ & & $1.15 \pm 0.15$ &
      $254/135 = 1.9$ \\
      && \hfill zphabs(rjcool-rjcool)) &
      $0.11 \pm 0.03$ & $0.54 \pm 0.03$ &
      $\dot{M} = 3.5 \pm 0.3$ \\

      \\
      Plume radius & 24694 &
      phabs(apec + apec) &
      $0.16 \pm 0.01$ & $1.05 \pm 0.03$ & $1.89 \pm 0.04$ &
      $0.97 \pm 0.07$ & $281/192 = 1.5$ \\

    \end{tabular}

    \caption{Results of spectral fits. Model names are shown in
        \textsc{xspec} notation.
        ${}^{*}$Fit with a 2-T model, with
        an additional component fixed at 4~keV and $0.4\Zsun$.
        ${}^\dagger$Intrinsically-absorbed cooling flow plus a plasma
        component fixed at 4~keV and $0.4\Zsun$. These are both
        absorbed by a fixed absorber, set to the galactic value of
        absorption $(0.081 \times 10^{22}\pcmsq)$. The temperature
        shown is the upper temperature of the cooling flow.
        $N_\mathrm{H}$ is the intrinsic cooling flow absorption.
        ${}^\ddagger$Like ($\dagger$), but subtracting another cooling
        flow component with the same normalisation and abundance, but
        different upper temperature as
        the first cooling flow. This simulates a cooling flow cooling
        from an upper temperature to a lower temperature (shown on the
        second line).
        ${}^\S$Like ($\dagger$), but with an additional
        \textsc{mekal} component to account for the local cluster
        emission. Its temperature is set to the upper
        temperature of the cooling flow, and its abundance is set to
        the same as the cooling flow.}
    
      \label{tab:regionfits}
    \end{center}
\end{table*}

\section{Imaging analysis}
\subsection{Radial profiles}
\label{sect:radial_prof}
In order to quantify the properties of the X-ray emitting gas as a
function of radius, we analysed the spectra of circular annuli around
the core of the cluster. This assumes that the properties of the gas
are largely symmetric, an assumption we will test in
\S\ref{sect:eastwest} and \S\ref{sect:adbin_fits}.  Spectra were
extracted, and appropriate background spectra, response and ancillary
matrices were created as in \S\ref{sect:spec_centre} for every
annulus.  We fitted each spectrum with 1-T and 2-T \textsc{mekal}
models, leaving metallicity and absorption as free parameters.  To
identify whether a 2-T fit was substantially better than a 1-T fit for
a particular annulus, we computed the $F$-statistic.  If the
probability of the observed value of $F$ (or a larger value) was less
than or equal to 0.01, we chose the 2-T fit for that annulus.  The
annuli were allowed to vary in width to ensure that there were
8000~counts in each spectrum (except for the outer annulus). Point
sources were excluded from the analysis.

\begin{figure}
  \begin{center}
    \includegraphics[angle=-90,width=0.99\columnwidth]{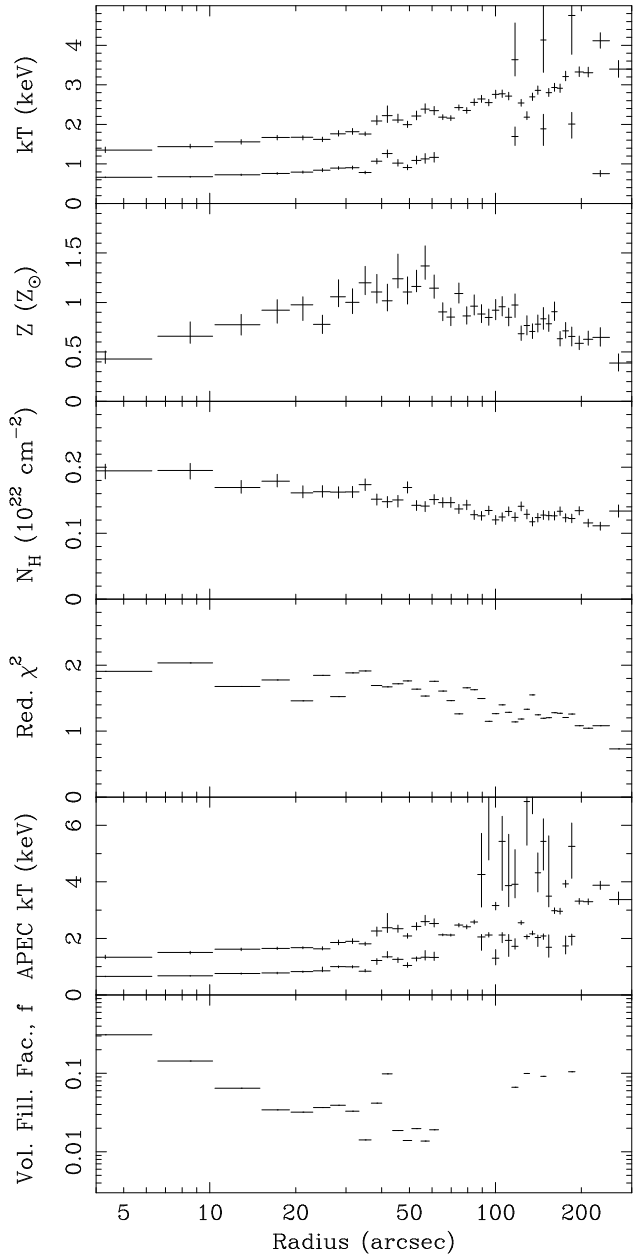}
  \end{center}
  \caption{Results of fits taken of spectra taken from radial
    annuli. Errors are 1-$\sigma$. A radius of 1 arcsec corresponds to
    2.04 detector coordinates, or a distance of 0.33\kpc. Radii at
    which there are two points have been fitted with a 2-T spectral
    model. The second-lowest panel shows the results of fitting with
    \textsc{apec} models, rather than \textsc{mekal} models. The
    lowest panel shows the volume filling factor, $f$, of the cool
    component (where appropriate).}
  \label{fig:radial}
\end{figure}

Fig. \ref{fig:radial} shows the radial profiles of temperature,
abundance, absorption and the reduced $\chi^2$ of the fits. The values
of the reduced $\chi^2$ show the fits are relatively poor in the
centre.  Two temperature points are given at the radii where the
$F$-statistic gave reason to do so, showing the temperatures of the
two temperature components.  The vertical error bars show the
1-$\sigma$ errors to the fit for each variable, whilst the horizontal
error bars mark the maximum and minimum radii of each annulus.

The temperature in the outer regions of the cluster is
$\sim~3.4$~\keV.  Towards the centre of the cluster, inside a radius
of $\sim 50$~arcsec (17~kpc), a 2-T model is required to fit the
spectra. The two temperatures drop to 1.35 and 0.66~keV at the centre
(consistent with \S\ref{sect:spec_centre}). The abundance rises from a
value of $\sim 0.4 \Zsun$ in the outer regions, to a maximum of $\sim
1.3 \Zsun$ at a radius of $\sim 45$~arcsec (15~kpc). It then drops in
the centre to $\sim 0.4 \Zsun$. The intrinsic absorption rises from a
value of $\sim 10^{21} \psqcm$ in the outer regions, to $\sim 2 \times
10^{21} \psqcm$ at the centre.  The lower-panel shows the temperature
results of fitting with 2-T and 1-T \textsc{apec} models, rather than
\textsc{mekal} models. The abundance and absorptions are very similar
to the \textsc{mekal} results, so we do not plot them here.

Using models with variable abundances (\textsc{vmekal} and
\textsc{vapec}), we find the relative abundances near the core are
more similar to those found in Type Ia supernovae rather than in Type
II supernovae (defined by ratios given in Dupke \& Arnaud 2001),
tieing together those elements which are produced by the same process
in each type of supernova. An even better fit is found by allowing the
abundances to be free: we find abundances (relative to solar), for an
annulus 35~arcsec away from the core using \textsc{vapec}, of O=0.5,
Mg=0.8, $\textrm{Ne}=2\times 10^{-6}$, Si=1.3, S=1.5, Ar=1.5, Ca=2.3,
Fe=1.1, and Ni=1.8.  \textsc{vmekal} gives different results: O=0.7,
Mg=0.2, Ne=0, Si=1.8, S=2.3, Ar=2.4, Ca=3.4, Fe=2.1, and Ni=2.1. The
level of Mg depends in detail on how the Fe~\textsc{l} emission is
modelled.

The volume filling fraction, $f$, of the cooler component in the 2-T
fits is given by
\begin{equation}
  f=\left[1+\frac{EM_2}{EM_1}
    \left(\frac{T_2}{T_1}\right)^2 \right]^{-1},
\end{equation}
where the emission measures and temperatures of the cooler and hotter
components are $EM_1, T_1$ and $EM_2, T_2$, respectively. This assumes
thermal pressure balance between the two components.  The radial
profile of $f$ (Fig. \ref{fig:radial}) shows that the cooler gas is
detected when it exceeds about one per cent of the volume; it exceeds
10 per cent within the inner 10 arcsec.

\begin{figure}
  \begin{center}
    \includegraphics[angle=-90,width=0.99\columnwidth]{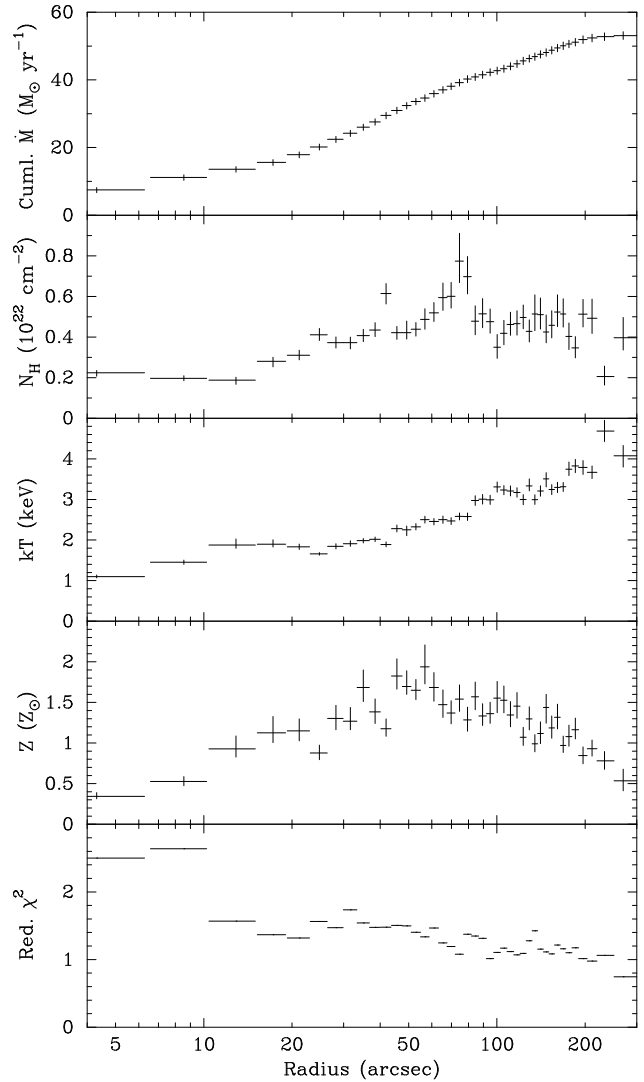}
  \end{center}
  \caption{Cumulative mass deposition profile, calculated by fitting
    the mass deposition rate in annuli, and summing from the centre
    outwards. Also shown are the intrinsic absorption of the cooling
    flow, its upper temperature, the abundance of the gas, and the
    reduced $\chi^2$ of the fit. The radiative cooling time of the gas
    at 300~arcsec is $\sim 10^{10}$~yr, decreasing inward to
    $<10^9$~yr within about 60~arcsec.}
  \label{fig:mdot_radial}
\end{figure}

We also fitted an absorbed cooling flow model to each annulus. We used
an intrinsically absorbed cooling flow (\textsc{rjcool}) plus a
\textsc{mekal} model, both absorbed by fixed galactic absorption. The
temperature and metallicity of the plasma model were fixed to the
upper temperature and abundance of the \textsc{rjcool} model.
Fig.~\ref{fig:mdot_radial} (top) shows a plot of the cumulative mass
deposition rate as a function of radius: the combined mass deposition
rate of all the annuli inside a particular radius. The plot rises from
a value of $\sim 7\Msunpyr$ in the central 6~arcsec, to $\sim 50
\Msunpyr$ at large radii (300~arcsec; 100~kpc), in agreement with
previous measurements for the mass deposition rate (Allen et al.
2001). Also shown in Fig. \ref{fig:mdot_radial} are the intrinsic
absorption of the cooling flow, its upper temperature, its
metallicity, and the reduced $\chi^2$ of each fit.

As with the spectrum of the core of the cluster, we fitted each
annulus with a CFMT model (the minimum temperature for each
annulus rises from 0.4~keV at the centre to 1~keV at a radius of
1~arcmin).  This model is only a significant improvement to the
standard CF model inside a radius of around $6.5$~arcsec
from the core.

We note that the level of absorption found is always higher than the
expected galactic value. This may, in part, be an ACIS-S calibration
issue. When fitting cooling flow spectra it may also characterize the
lack of low temperature emission lines, such as are absent from
\emph{XMM-Newton}/RGS spectra of cooling flows. We do not have the
spectral resolution or sensitivity to detect absorption edges which
would demonstrate the presence of intrinsic absorption.

\subsection{East-west anisotropy}
\label{sect:eastwest}
\begin{figure}
  \centering
  \includegraphics[width=0.99\columnwidth]{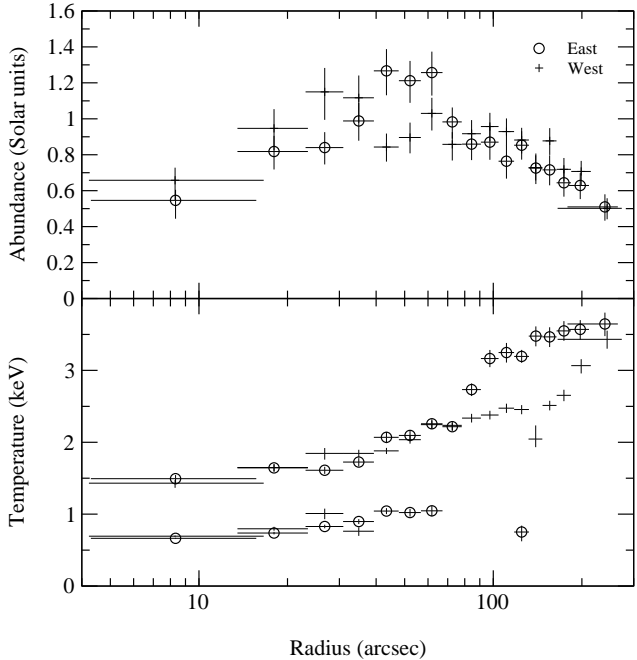}
  \caption{Radial abundance and temperature profiles of the east and
    west of the cluster.}
  \label{fig:binlr}
\end{figure}

The east side of the cluster contains the plume-like feature. To
investigate whether there is significant difference between the east
and west sides of the cluster, we analysed annuli cut in a north-south
direction. Fig.~\ref{fig:binlr} shows the two temperature and
abundance profiles for the cluster. The annuli were chosen to contain
20000 counts between 0.5 and 7~keV, and then were split into east and
west parts. 1-T and 2-T \textsc{mekal} models were fitted using the
same procedure as in \S\ref{sect:radial_prof}, and an $F$-test was
used to decide whether the 2-T model was a significant improvement
over the 1-T model.

The plot shows significant differences in abundance between the two
sides at radii of $\sim 40$--50~arcsec. There are also sharp outward
rises in temperature in the east and west, at radii of $\sim 70$ and
160~arcsec, respectively. We will discuss these further in the next
section. Between these radii, there are temperature differences
between the two halves of the cluster. These radii match the region in
\textsc{mekal} and \textsc{apec} radial plots where two temperature
components are often preferred (Fig.~\ref{fig:radial}).  The
anisotropy shows that more detailed analysis is required to map out
the structure of the cluster.

\subsection{Fits to adaptively binned areas}
\label{sect:adbin_fits}
\begin{figure*}
  \begin{tabular}{llll}
    \raisebox{6mm}{(a)} & \includegraphics[width=0.39\textwidth]{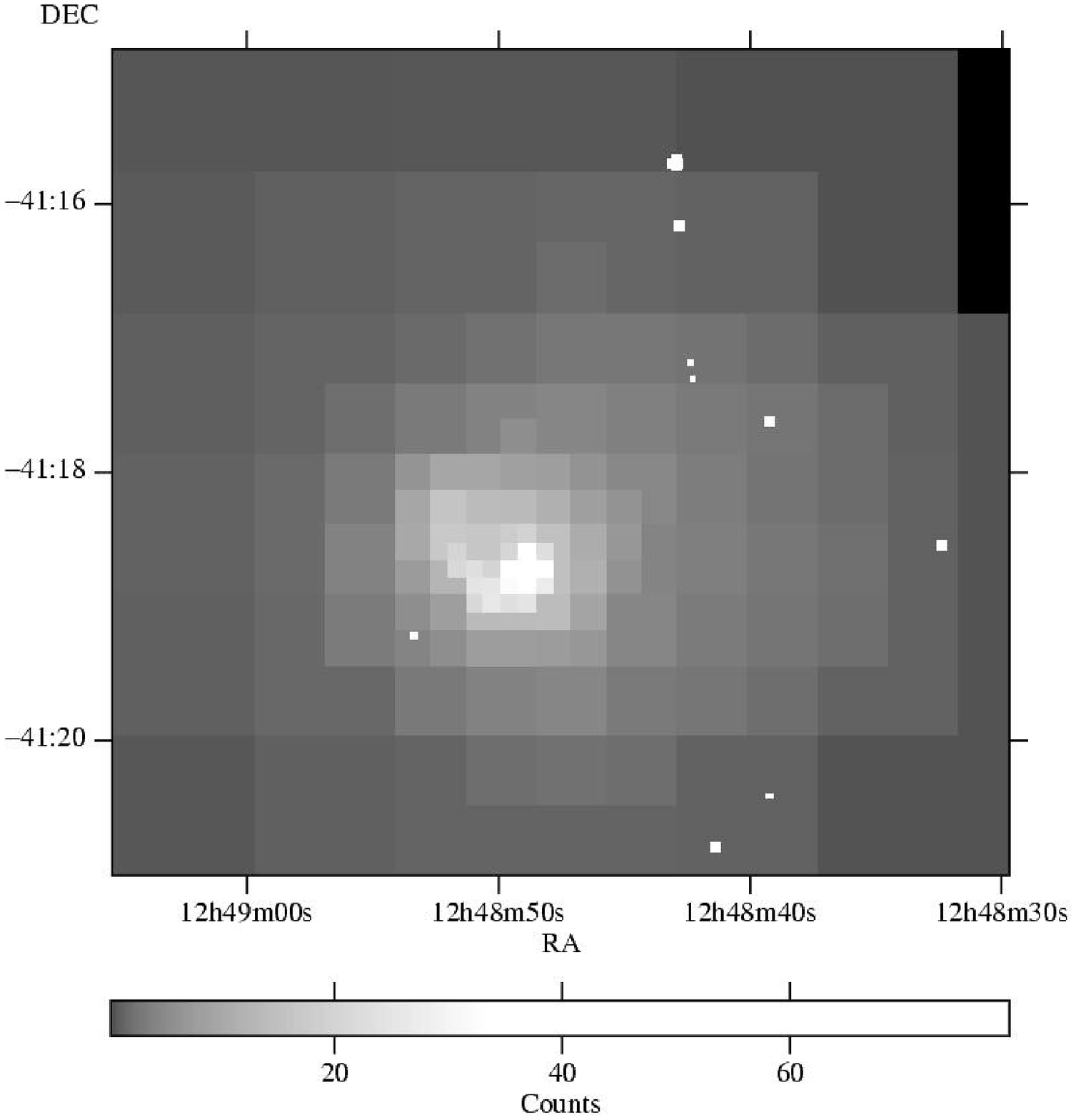} &
    \raisebox{6mm}{(b)} & \includegraphics[width=0.39\textwidth]{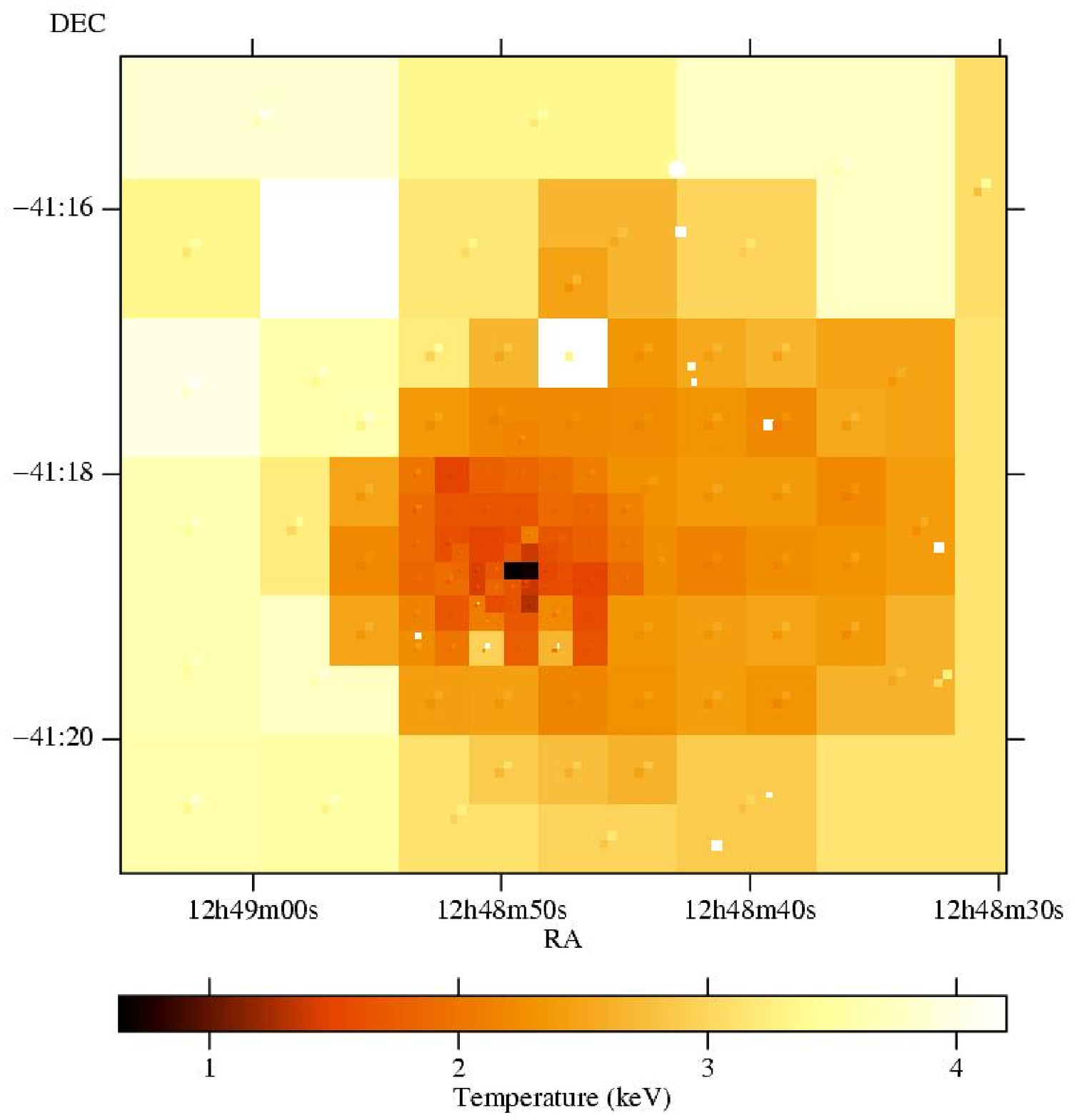} \\
    \raisebox{6mm}{(c)} & \includegraphics[width=0.39\textwidth]{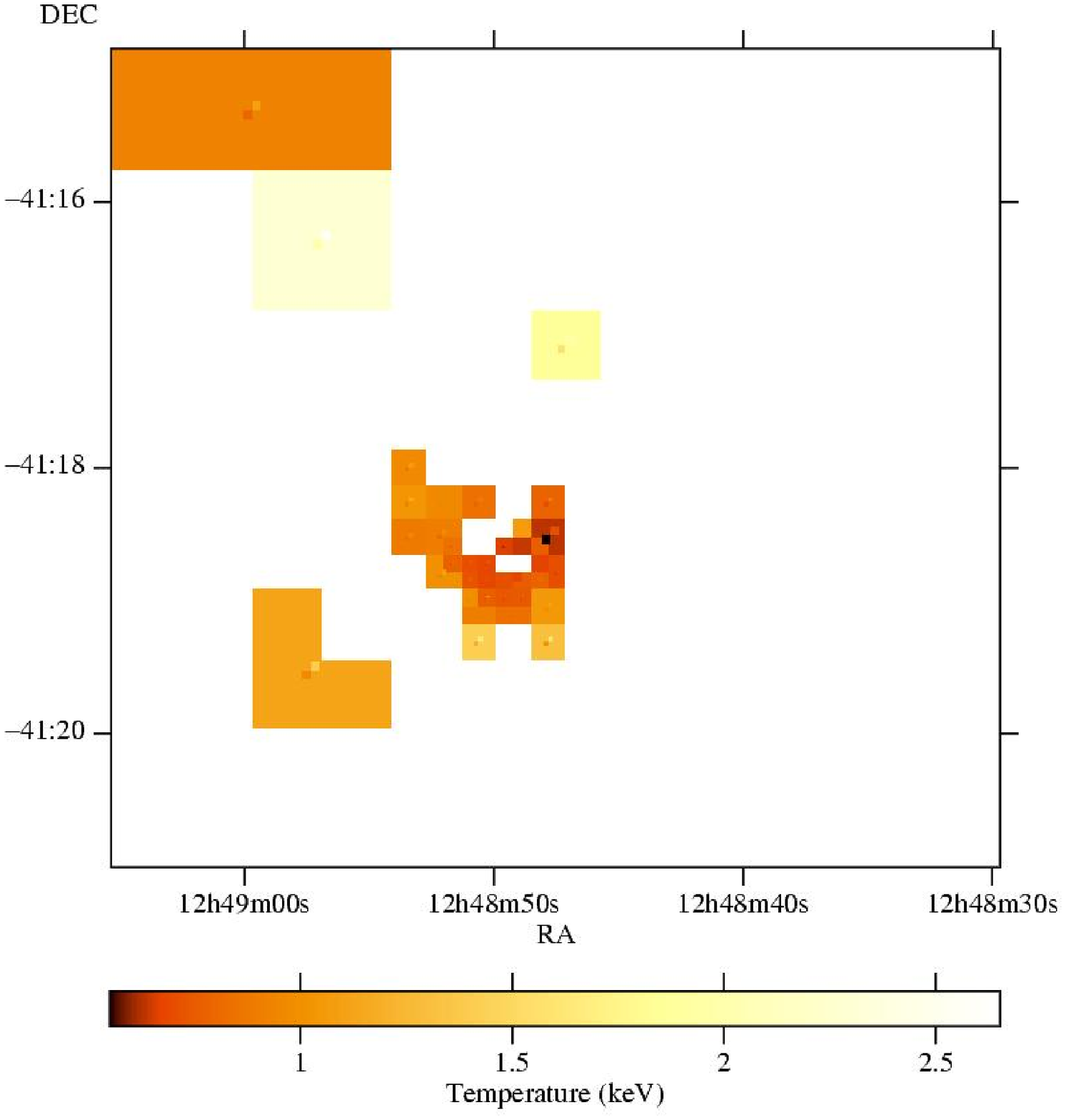} &
    \raisebox{6mm}{(d)} & \includegraphics[width=0.39\textwidth]{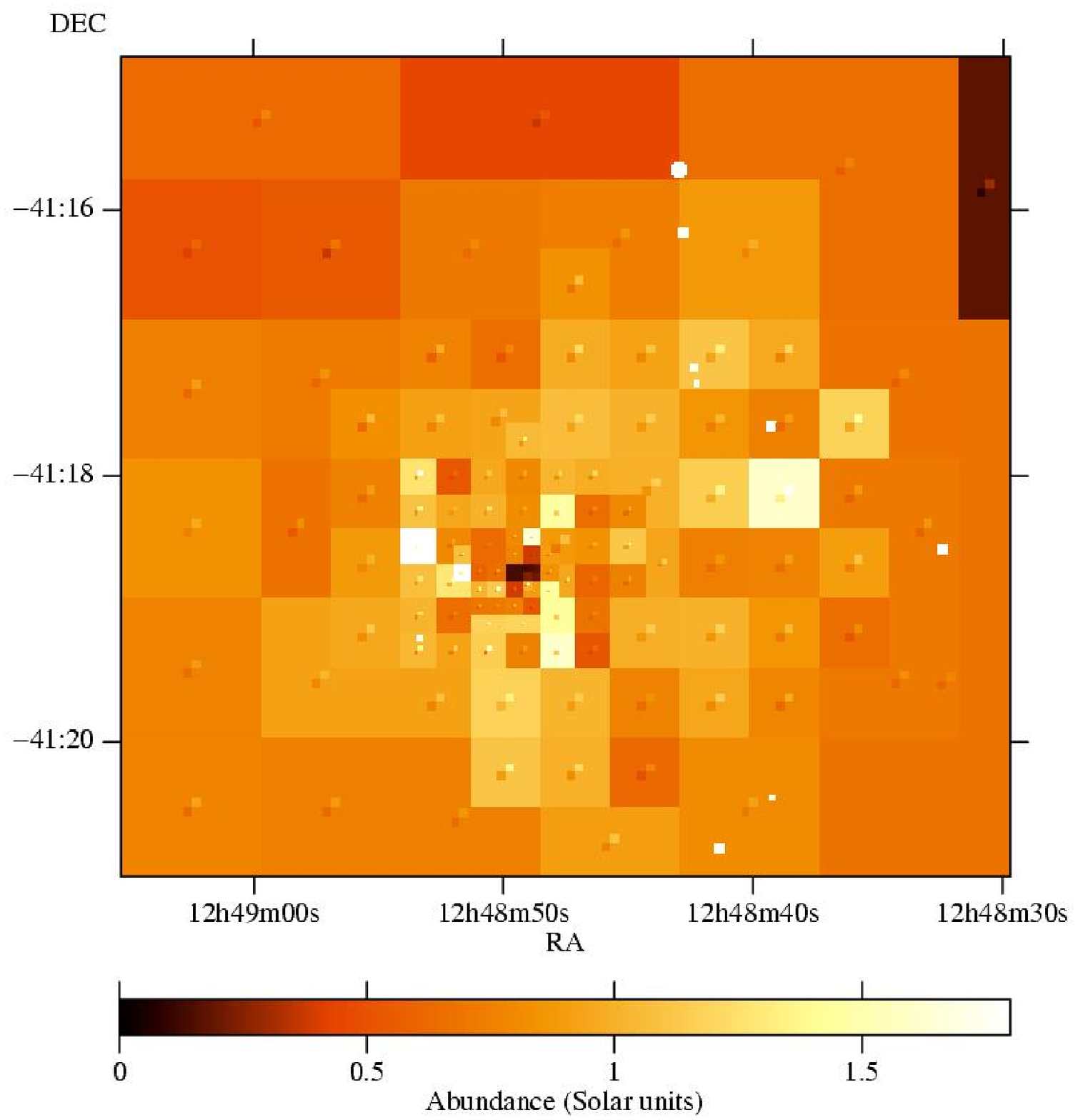} \\
    \raisebox{6mm}{(e)} & \includegraphics[width=0.39\textwidth]{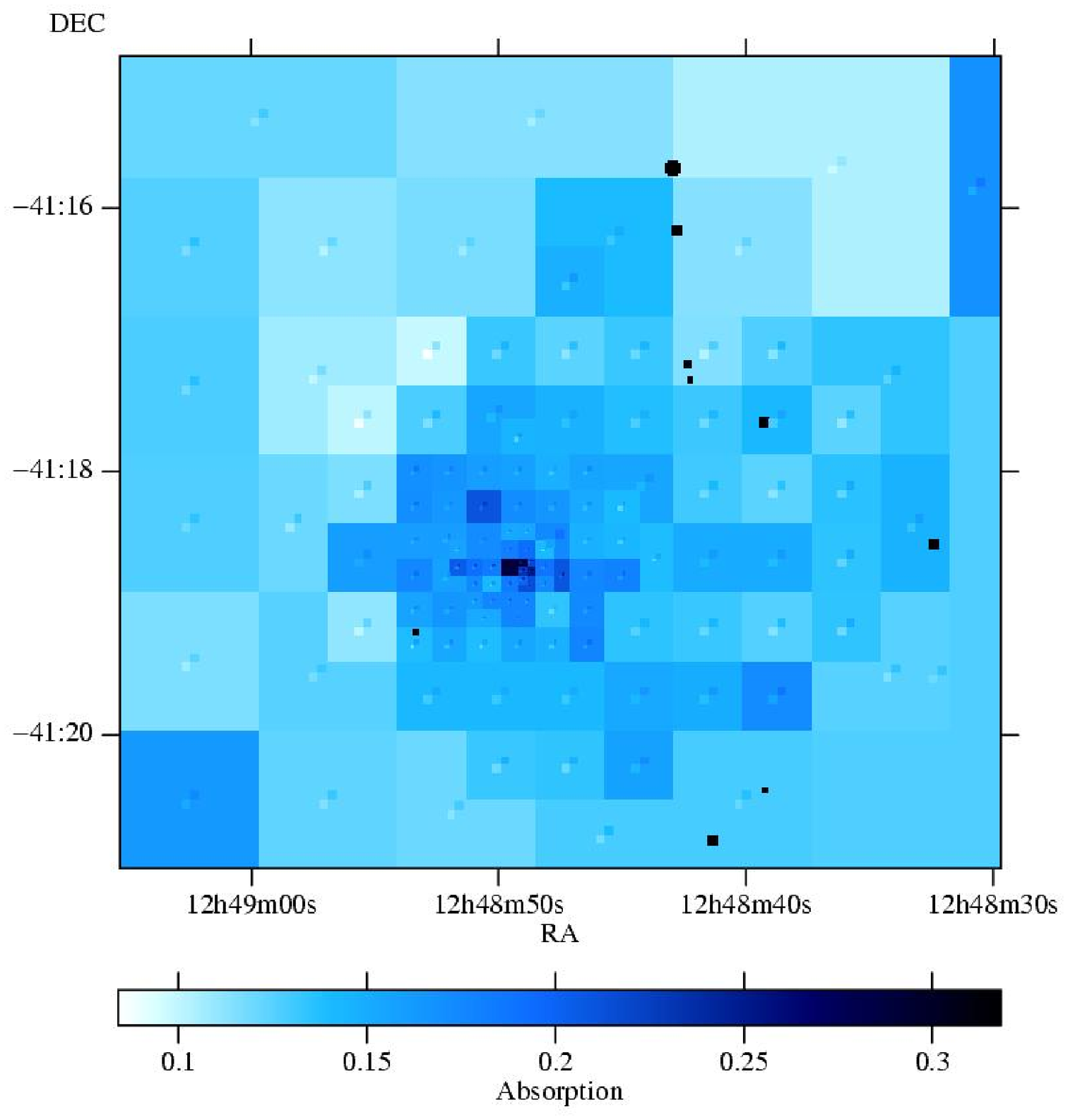} &
    \raisebox{6mm}{(f)} & \includegraphics[width=0.39\textwidth]{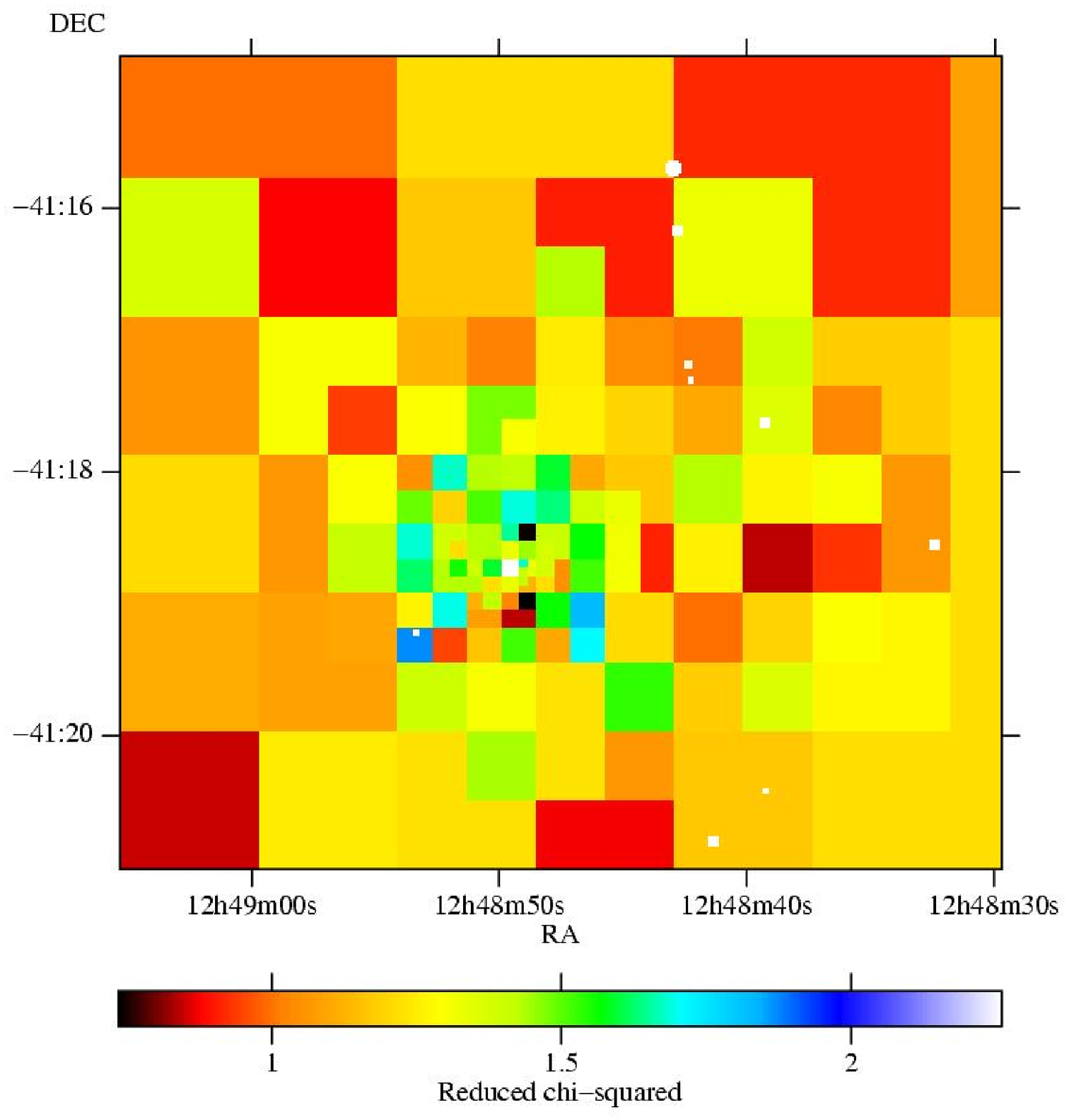} \\
  \end{tabular}
  \caption{(a) Adaptively binned image of the cluster, counts are average
    values per arcsecond pixel.
    (b) Upper-temperature results (for 2-T fits) or the 1-T fit
    results. The diagonal boxes inside each bin
    indicate the bounds of the 1-$\sigma$ error of the temperature.
    The white spots are excluded point-sources.
    (c) Lower-temperature fit results for the 2-T fits.
    White regions are where 1-T fits were used.
    (d) Abundance results for fits.
    (e) Absorption results of fits.
    (f) Reduced $\chi^{2}$ of the fits.}
  \label{fig:adbin}
\end{figure*}

We adaptively binned (Sanders \& Fabian 2001) the central 3 arcmin of
the cluster, using a fractional error threshold of 0.03. Adaptive
binning uses a bin size which changes to ensure a minimum number of
counts in each bin. A threshold of 0.03 gives 1100 counts per bin
(except in certain cases).  Fig. \ref{fig:adbin}(a) shows the
adaptively binned image of the cluster. Point-sources were excluded
from the analysis and are marked as white areas. The core of the
cluster is shown to the south-east of the centre of the image.

We extracted spectra from each bin, and generated appropriate
background spectra, response and ancillary matrices, as in
\S\ref{sect:spec_centre}. We then fitted each bin with 1-T and 2-T
\textsc{mekal} models, finally using an $F$-test to decide whether the
2-T fit was a significant improvement over the 1-T fit.  The results
for the two types of fit were then combined to create images showing
the spatial distribution for each fitted parameter.  Fig.
\ref{fig:adbin}(b) shows either the fitted temperature from the 1-T
fit, or the upper temperature of the 2-T fit. Fig.  \ref{fig:adbin}(c)
shows the lower temperature of the 2-T fit, where appropriate.  Figs.
\ref{fig:adbin}(d) and (e) show the fitted abundance and absorption
levels from the relevant fit.  Fig.  \ref{fig:adbin}(f) shows the
appropriate reduced $\chi^2$ value.

The value of a parameter in the images is indicated by a colour scale.
Subtracted point sources are marked by the maximum value.  We plot the
1-$\sigma$ errors to the fit as a pair of diagonal boxes in the centre
of each bin. The top-right box shows the upper bound, and the
bottom-left shows the lower bound. Masked-out point sources are shown
as areas coloured with the maximum value.

2-T models are required to fit the spectra in the inner 30~arcsec.
The 2-T temperature results in the centre ($\sim0.7$ and 1.2~keV) are
cooler than the 1-T results in the outer regions ($\sim 3\keV$).  The
abundance of the gas in the outer regions has a similar value to that
in the core ($\sim 0.5\Zsun$), but there is a rise in the metallicity
about 1~arcmin away from the core to $\sim 1.3 \Zsun$, as in
\S\ref{sect:radial_prof}.

The images show the properties of the emitting gas are not spherically
symmetric. The temperature sharply increases at between 70 and 160
arcsec radius, depending on position angle (Fig.~\ref{fig:binlr}).
Such features have been seen in other clusters, and are termed cold
fronts (Markevitch et al. 2000). The most striking front is to the
north-east of the core in the Centaurus cluster, where the emission
suddenly drops.  There is an extensive cooler region to the west of
the core ($\sim 2.4\keV$; Fig.
\ref{fig:adbin}(b)). This area has a higher abundance than an
equivalent region to the east (Fig.
\ref{fig:adbin}(d)). In addition, there is more absorption (Fig.
\ref{fig:adbin}(e)) to the west.

In the innermost core of the cluster (at a radius of less than
4~arcsec) there are a few bins for which a 1-T fit is preferred using
an $F$-test (probabilities for $F$ of $\sim 0.03$). Owing to this,
they have much lower abundances than the surrounding bins. If their
spectra are combined, a 2-T fit is easily preferred, and the results
of the fit closely match the values of the surrounding bins.  This is
likely to be due to the difficulties of detecting two low ($\sim
1$~keV) temperatures close together. More counts at that temperature
appear to be required to detect two temperatures.

\begin{figure}
  \includegraphics[width=0.99\columnwidth]{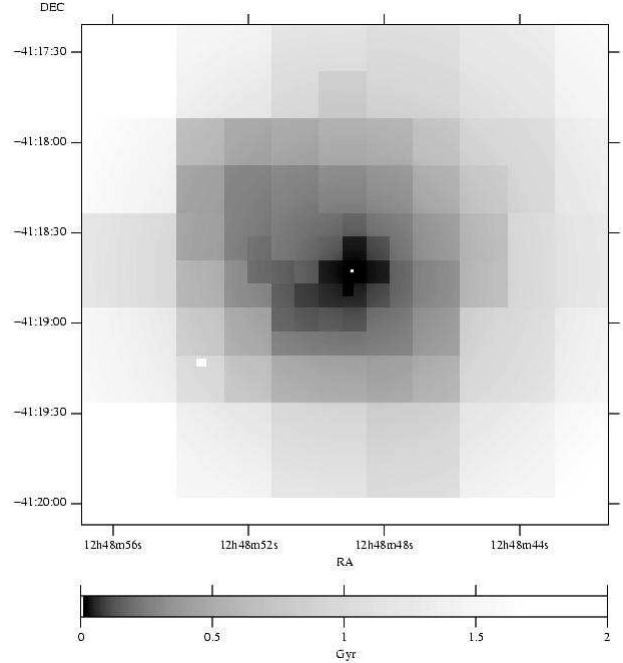}
  \caption{Mean radiative cooling time of the gas, estimated from the X-ray
    surface brightness (see Fabian et al. 2000). The mean cooling time
    is less than $5 \times 10^8$~yr inside a radius of 45~arcsec,
    below $10^8$~yr inside 10~arcsec (3~kpc) radius, and falls to $3
    \times 10^7$~yr within 5~arcsec of the centre.}
 \label{fig:tcool}  
\end{figure}

\begin{figure}
  \begin{center}
    \includegraphics[width=0.95\columnwidth]{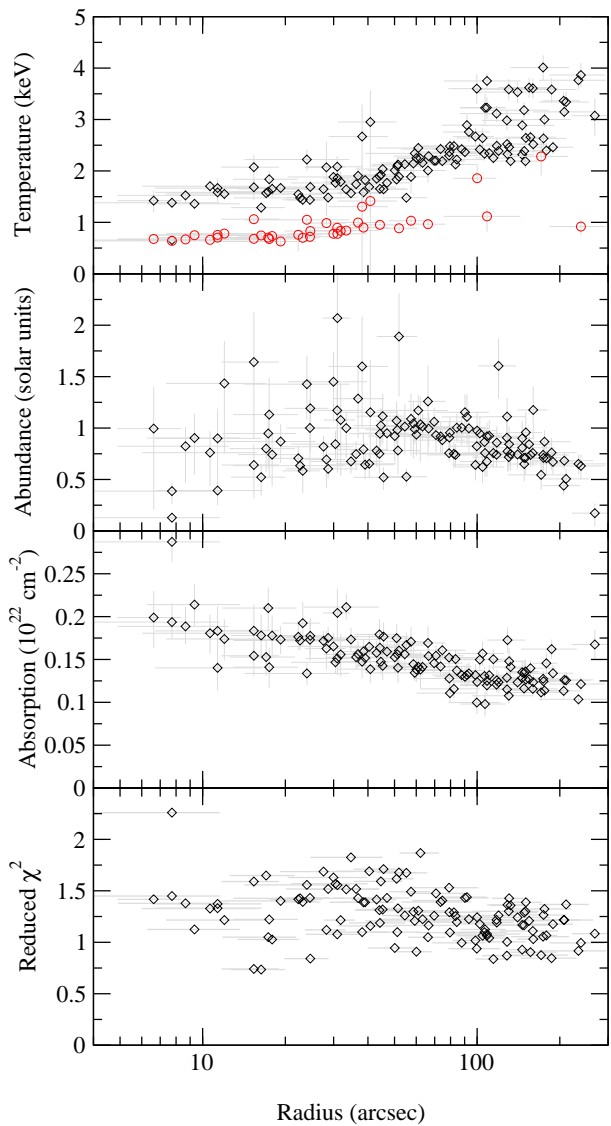}
  \end{center}
  \caption{Distribution of the fitted values for the bins in
    Fig.~\ref{fig:adbin} as a function of radius. The radius of a bin
    is the mean displacement from the centre of the cluster, and the
    error bars indicate the maximum and minimum radius of each pixels
    in the bin. The errors on the value are the 1-$\sigma$ fitted
    errors also shown in Fig.~\ref{fig:adbin}. Two points are plotted
    in the temperature plot for bins fitted with a 2-T model. Bins at
    a radius of less than 4~arcsec are excluded, due to an incorrect
    fitted model (see text).}
  \label{fig:bin_plot}
\end{figure}

In order to examine whether there is significant non-radial variation
in the bins, the fitted parameters (with 1-$\sigma$ errors) to each
bin as a function of the radius of the bin are plotted in
Fig.~\ref{fig:bin_plot}.  The radius plotted is the mean radius of the
pixels of the bin. The plot brings out the features in
Fig.~\ref{fig:radial}.  There is some significant variation in
metallicity between radii of 7--50~arcsec. Examining those bins in the
east and those to the west, we find that most of the variation in the
abundance is to the west (the opposite side to the plume). The
metallicity to the west is scattered between 0.5 and 1.7~\Zsun at
radii of 7--50~arcsec. The profile to the east is much smoother,
except for an obvious high abundance point (Fig.~\ref{fig:adbin}(d),
roughly 7~arcsec from the core). The east-west difference in
metallicity and temperature is highlighted by Fig.~\ref{fig:binlr},
which shows east-side and west-side profiles of the cluster.

Given that the cluster contains a strong radio source (Taylor et al.
2001) there may be significant non-thermal emission in the X-ray band.
Any non-thermal emission will increase the continuum of the X-ray
emission and lower the apparent metallicity.  We tested for the
presence of non-thermal emission by fitting the spectrum in each bin
with a 1-T plus power-law model. We then used an $F$-test to decide
whether adding the power-law component made a significant improvement
to the fit (using the same criterion as \S\ref{sect:radial_prof}) for
each bin.

\begin{figure}
  \begin{center}
    \includegraphics[width=0.95\columnwidth]{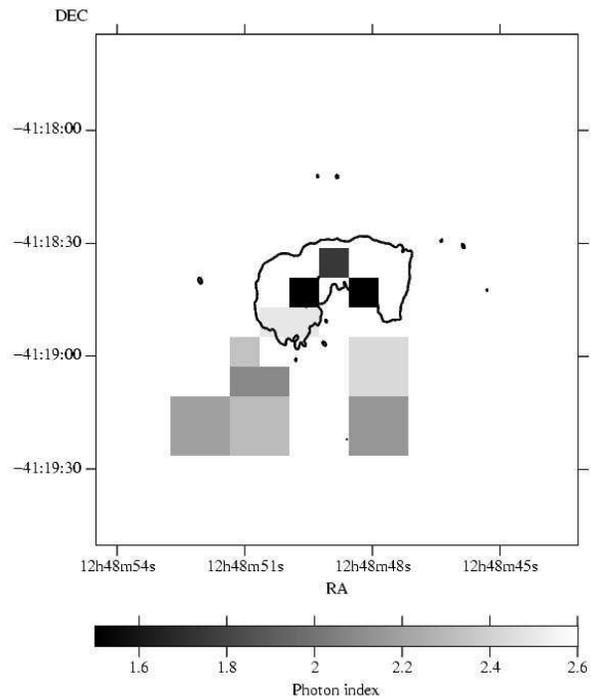}
  \end{center}
  \caption{Regions in the centre of the cluster where a power-law
    plus 1-T is an improvement over a 1-T fit.  White areas indicate
    where the $F$-test gives no significant improvement over a 1-T
    fit. The other regions are shaded to show the photon index of the
    best-fitting power-law.  The photon index was contrained to lie
    between 1.5 and 2.5. The contour marks where the radio emission is
    0.2~mJy~beam$^{-1}$ at 4.8~GHz, taken from the data of Taylor et
    al.  (2001).}
  \label{fig:powerlaw_radio}
\end{figure}

We show in Fig. \ref{fig:powerlaw_radio} the areas where the power-law
component is suggested by the $F$-test. Interestingly the morphology
of the radio source approximately matches those areas in the centre
where the power-law component is a significant improvement. We also
find that those areas 1~arcmin south of the core where the power-law
is preferred match the low-intensity regions of the radio source (see
Taylor et al. 2001 for detailed maps of the radio source morphology).
However, only a small fraction of the emission from the core of the
cluster appears to have a significant non-thermal component, and the
peak in the abundance still exists if we exclude those regions where
it may be important.  Therefore the peak in abundance 15~kpc from the
core is not an effect of non-thermal emission.

\section{Discussion}
The most interesting conclusion of our analysis is that the
metallicity of the cluster peaks at $\sim 1.3 \Zsun$ (or $1.8\Zsun$
with a cooling flow model) at a radius of $\sim 15$~kpc, declining to
$0.4 \Zsun$ at the centre. Previous observations (Fukazawa et al.
1994; Ikebe et al. 1998) of the cluster did not have high enough
spatial resolution to resolve this feature, showing only an increasing
gradient in metallicity to the centre.  B\"ohringer et al.~(2001)
found using \emph{XMM-Newton} data that the abundances in the Virgo
Cluster peak $\sim 1$~arcmin from the centre.  Molendi \& Gastaldello
(2001) later saw this to be an effect of using a 1-T fit where more
temperature components were required; within 1~arcmin from the centre
the actual metal profile is flat.

\begin{figure}
  \centering
  \includegraphics[angle=-90, width=0.95\columnwidth]{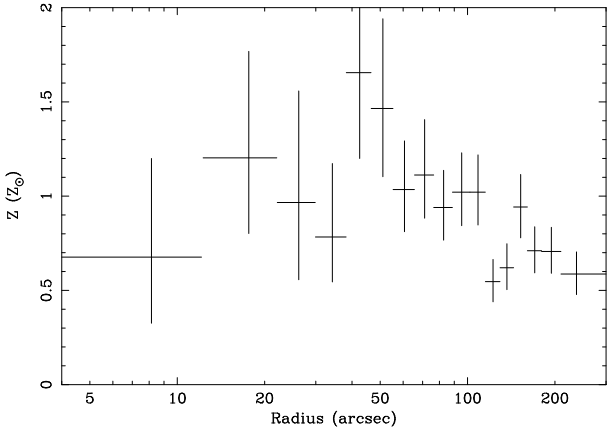}
  \caption{Radial metallicity plot generated by fitting spectra
    between 3 to 7~keV with a 1-T model (above 3~keV only one
    temperature component is necessary). Each radial bin contains
    $\sim 20000$~counts from 0.5--7~keV.}
  \label{fig:zhighen_radial}
\end{figure}

The abundance gradient in Centaurus was modelled by Reisenegger,
Miralda-Escud\'e \& Waxman (1996), but no decrease was expected.  We
still get a peak in abundance just fitting spectra between 3 and 7~keV
where Fe~\textsc{k} emission occurs (Fig.~\ref{fig:zhighen_radial}),
so the peak is not an artefact of the Fe~\textsc{l} complex. Resonant
absorption is a possible mechanism for the metallicity decline in the
core of the cluster. However, since we see the abundance peak in the
Fe~\textsc{k} lines too, resonant absorption is unlikely.

In the future, it will be important to study the chemical composition
of the plasma, and handle projection effects fully. We also would like
to understand why the statistical quality of the fits in to the
spectra in the centre are poor, and how to improve them. These effects
are probably due to the high metallicity of the core: a feature which
is particular to this cluster. We would like to draw attention to the
fact that there are differences in the qualities of fits using
different plasma codes.  The fits to the adaptively-binned image
spectra (Fig.~\ref{fig:adbin}(c)) suggest that the 2-T fits are
required in specific regions, rather than at specific radii.

Using a CF model produces a radial abundance profile with a similar
shape, but with a higher peak metallicity, than that found using the
2-T model, indicating that the peak is not a model-dependent effect.
The upper-temperature of the CF fit also matches the upper-temperature
of the 2-T fit. In general, excluding the very central bins, the CF
model gives a smaller reduced $\chi^2$ than the 2-T model.

It is difficult to conclude in general which model gives the better
fits to the regions of the cluster. When the data are binned in
annuli, a standard cooling flow model (CF) is a significant
improvement (measured using an $F$-test with a maximum probability of
0.01) over a 1-T fit for nearly every annulus. 2-T and CFMT models are
only a significant improvement over a 1-T model within about 60~arcsec
of the cluster core.  A 2-T model or CFMT model are a significant
improvement to the CF model within the inner $\sim 7$~arcsec.  A CF
model probably fits the spectra of the annuli better as the cluster is
not radially symmetric, so there are a number of temperatures at each
radius, which the CF model includes.

The comparison gives different conclusions when fitting the regions in
the adaptively binned image. A 2-T, CF or a CFMT model is a
significant improvement (with the same $F$-test criterion) over the
1-T model inside much of the inner 40-50~arcsec. The area in the
central arcmin for which the CF model is an improvement over the 1-T
model is roughly $2/3$ that for which the CFMT and 2-T models are an
improvement. The CFMT and 2-T models show a significant improvement in
the fit over the CF model for areas about 50~arcsec away for the
nucleus, and some of the area inside. Fig.~\ref{fig:tcool} shows that
the radiative cooling time of the gas is less than a Gyr within those
radii. Outside the core, there are some areas which show a significant
improvement using a CF, 2-T or CFMT model over a 1-T model. It may be
that there are local variations in temperature in these regions, or
this could be a statistical effect. Using a larger bin size, with the
corresponding increase in counts in the spectra, still results in
areas outside the core which show a significant improvement using a
2-T or CFMT model over a 1-T model, although using a larger bin size
includes more gas, which is probably at different temperatures. Note
that if there is a cooling flow, then the models used here do not
incorporate the gravitational work done on the gas in the flow.
Therefore no firm statement can be made on the level of the mass
deposition rate. The data support a rate of several tens \Msunpyr.

It is difficult to speculate on the cause of the plume-like feature in
the core of the cluster. The galaxy velocity data cannot rule out a
100-200~\kmps line of sight motion of NGC~4696 in Cen~30 similar to
that of the cD in A1795 (Fabian et al. 2001b). The real velocity
could of course be larger and there may be significant motion of the
intracluster gas within its potential well.  The plume does, however,
touch the east-most extent of the radio lobes (Taylor et al. 2001),
indicating there may be some interaction between the lobes and the
plume. However, the direction of the plume is not aligned with the
radio lobes. The plume has a higher metallicity than the core,
suggesting the cooling flow is depositing enriched material in this
area.

Sparks, Macchetto \& Golombek (1989) and de Jong et al. (1990) have
proposed that the cooler central gas around NGC~4696 originates from an
infallen dust-rich dwarf galaxy, which also accounts for the dust
lane. The fact that the metallicity of the plume matches its
surroundings argues against this stripping hypothesis (as mentioned in
\S\ref{sect:spec_plume}). Moreover, cool gas is found in the majority
of clusters.

As in M87, the plume could be due to a radio bubble dragging out
cooler gas (B\"ohringer et al. 1995; Churazov et al. 2001).  The `cold
front' at the end of the plume could in fact be the edge of the
bubble. However, we have shown that the metallicity of the plume is
similar to the surrounding material, and higher than the central gas,
which argues against that gas having originated close to the central
radio source.  Of course, the drop in metallicity within the inner
20~kpc (60~arcsec) could be due to bubbles transporting higher
metallicity gas to larger radii (Churazov et al. 2001; Br\"uggen \&
Kaiser 2001). However, it is not then clear where the present
lower-abundance gas has come from.  More detailed models of the
convection and exchange of gas are required, but at first inspection,
such models are unlikely to reproduce the abundance profile seen here.

Our main conclusion is that the steep abundance gradient in this
cluster does not persist within 15~kpc, but decreases inward to a
value similar to that seen at large radii. This may be a feature of
radiative cooling of the gas, if composed of small metal-rich clumps
in a metal-poor surrounding environment (Morris \& Fabian, in
preparation).  \emph{The presence of an abundance gradient means that
  there has not been significant widespread convection or mixing
  extending beyond 15~kpc driven by the nucleus of NGC~4696}. The
spectra within $\sim 60$~kpc also indicate that the temperature
structure of the gas has at least two phases, both decreasing towards
the centre.  The present data are not capable of testing for the
presence of further phases.

\section*{Acknowledgements}
We thank the referee, Silvano Molendi, for helpful comments.  ACF and
JSS would like to thank the Royal Society and PPARC for support,
respectively.

\end{document}